# Automating Knowledge-Driven Model Recommendation: Methodology, Evaluation, and Key Challenges

Adam A. Butchy, Cheryl A. Telmer, and Natasa Miskov-Zivanov

**Abstract**—There is significant interest in using existing repositories of biological entities, relationships, and models to automate biological model assembly and extension. Current methods aggregate human-curated biological information into executable, simulatable models, but these models do not resemble human curated models and do not recapitulate experimental results. Here, we outline the process of automated model assembly and extension, while demonstrating it on both synthetic models and human-curated models of biological signaling networks. We begin with an iterative, greedy, and combinatoric approach to automated assembly and demonstrate the key difficulties inherent to contextless assembly. We publicly release the software used in this paper to enable further exploration of this problem.

---

## 1 INTRODUCTION

Computational approaches to modeling large complex systems standardize the representation of knowledge, while simulation of computational models illuminates the dynamics of systems, allowing for discoveries and theoretical advances [1]. Due to the complexity and redundancy of biological systems, computational models are difficult and laborious to create and update. There are two main approaches to modeling these systems, bottom-up and top-down [2]. In a bottom-up approach, known molecular interactions are assembled into a model to help explain the system's behavior and predict how the system will respond to new stimuli or inputs. This method has been used extensively by biologists, biochemists, and molecular biologists to manually create models based on the interactions within cells involved in signaling that are supported by scientific literature. In a top-down approach, experimental data—usually collected with high-throughput methods—is used to infer correlations between element behavior and determine causal relationships. Top-down approaches employ many different methods such as Bayesian Inference [3], ANOVA calculations [4], and Fuzzy Logic [5]. In both the mechanistic bottom-up approach and the data-driven top-down approach, the model is used to predict the behavior of individual elements in the network [6, 7]. Recently, there has been a push to integrate the two methods, using experimental data to inform the bottom-up approach, and incorporating prior knowledge into the top-down approach to reduce the number of potential models [8-12]. Despite these hybrid approaches, this problem remains a combinatoric one, with large, complex systems being prohibitively difficult to investigate and model manually.

It is a direct result of these factors that system and computational biologists have endeavored to automate the process of model creation and extension. To automatically create models, information can be extracted from literature, queried from databases, or taken from existing pathways and models. Public databases such as Reactome [13], MetaCyc [14], OmniPath [15], and STRING [16] offer easy access to millions of interactions. Additionally, there exist a number of model databases with published models that are publicly available such as The Nature Pathway Interaction Database [17], WikiPathways [18], BioModels [19], the Cell Collective [20], and KEGG pathways [21]. These databases contain highly targeted, curated published and unpublished models which are created for specific biological context and may not be generalizable to explain other phenomena. When new interactions are discovered, and described in a scientific publication, state-of-the-art machine reading engines such as REACH [22], TRIPS [23], and EVEX [24] can extract them, together with other relevant information. These automated readers are able to extract tens of thousands of biological entity interactions from hundreds of papers in a few hours, and produce a machine-readable, structured output [22]. Despite this abundance of available interactions, there is still no efficient way to assemble them into accurate models that correctly reflect the system under investigation and the same biological context and recapitulate the observed experimental behavior.

Recently, a few tools, such as Path2Models [25] and IN-DRA [26, 27], have been created to help modelers collect biological interactions, assemble a model, and perform

---

- *A.A. Butchy is with the Department of Bioengineering, University of Pittsburgh, Pittsburgh, PA 15213. E-mail: adam.butchy@pitt.edu.*
- *C.A. Telmer is with the Department of Biological Sciences, Carnegie Mellon University, Pittsburgh, PA 15213. E-mail: ctelmer@cmu.edu.*
- *N. Miskov-Zivanov is with the Departments of Electrical and Computer Engineering, Bioengineering, and Computational Biology, University of Pittsburgh, Pittsburgh, PA 15213. E-mail: nmzivanov@pitt.edu.*

simulations. These tools assemble quantitative and qualitative models using available pathway information; however, the quality of the assembled models is dependent upon the modeling approach, and the granularity of the information they are given. These techniques rely on accurate information, and their performance suffers when the interaction information is incomplete, from a different biological context, or erroneous. Other methods have been proposed to automatically expand, test, and select the best model, with respect to a given performance metric. These approaches integrate stochastic model simulations with statistical model checking only [28], or also incorporating Markov clustering [29], or genetic algorithm [30], and therefore have different strengths and weaknesses. The Markov clustering approach to model extension is well suited for the combinatorial explosion in the number of possible model extensions while the genetic algorithm approach is overwhelmed by large number of extensions. Markov clustering prioritizes strongly connected components at the expense of interactions involving nodes of low degree. The genetic algorithm explores the effect of single extensions distributed throughout the network.

In this work, we examine the complexities inherent to automatic model assembly and extension. We use two novel algorithms, Breadth First Addition (BFA) and Depth First Addition (DFA), which utilize the same principles as the breadth-first search and depth-first search algorithms in network studies [31] to illustrate the key limitations of iterative model assembly and extension. In contrast to previous work [28-30], these methods not only represent a new approach to bottom-up model assembly but are also used to demonstrate the existence of key biological properties which hinder automated modeling of biological systems. We demonstrate these properties using both synthetic networks, Erdös-Rényi random networks (ER) [32] and Barabási-Albert scale-free networks (BA) [33], as well as two published expert curated and validated models, a T cell large granular lymphocyte (TLGL) leukemia model [34], and a model of naïve Tcell differentiation (Tcell) [35]. By using different network structures, we are able to more comprehensively explore automated model assembly and identify the main difficulties with the BFA and DFA approaches.

## 2 METHODS

### 2.1 Discrete Models and Simulations

The underlying structure of models that we study here is a network $G(V, E)$, where $V$ is a set of nodes (model elements), and $E$ is a set of directed edges (regulatory influences between elements). A few toy examples of such networks are shown in Figure 1 (A). Model elements usually represent proteins, genes, chemicals, or biological processes. For each model element $v_i \in V$ ($i = 1..N$, where $N = |V|$), we define an update rule $v_i = f_{v_i}(v_1, v_2, ..., v_N)$, which can either be a constant (input nodes in network $G$) or it can depend on a subset of elements from $V$. In the latter case, for each element $v_i$ this subset is often referred to as an influence set for $v_i$ and it consists of its positive (activating) and negative (inhibiting) regulators. Positive regulators of $v_i$ comprise set $V_{pos}^i$ and are represented with regular arrowheads in Figure 1 (A). Negative regulators of $v_i$ comprise set $V_{neg}^i$ and are represented with blunt arrowheads in Figure 1 (A).

The high throughput retrieval of interaction information from literature typically only includes knowledge of the sign of influence (positive or negative) and rarely additional information about relationships between regulators. In such cases, logic functions and elements with two levels, 0 (low) and 1 (high), have been found most suitable. To broaden the application beyond just Boolean functions to other cases where interactions were enriched either through manual curation or more specific information retrieval, we will assume that each element $v_i$ can have $L_i$ number of discrete levels. While the choice of function does not affect the main algorithms described in Section 2.2, in order to simulate models, and closely approximate different functions, including Boolean, we adopted the common approach that computes a (weighted) sum of regulator values to determine element update values. The general form of this function is:

$$g_{v_i} = f_{v_i}(v_1, v_2, ..., v_N) = \sum_{v_j \in V_{pos}^i} w_j v_j - \sum_{v_k \in V_{neg}^i} w_k v_k \quad (1)$$

The weighting factors $w_j$ and $w_k$ can be used to account for different influence strengths for regulators. To remain within boundaries of the allowed levels for element $v_i$ (0..$L_i - 1$), the function $g_{v_i}$ is then used to determine a suitable increment/decrement for $v_i$, $\delta_{v_i} = f(g_{v_i})$, such that:

$$v_{i,next} = \begin{cases} 0 & v_i + \delta_{v_i} \leq 0 \\ v_i + \delta_{v_i} & 0 < v_i + \delta_{v_i} < L_i - 1 \\ L_i - 1 & v_i + \delta_{v_i} \geq L_i - 1 \end{cases} \quad (2)$$

Together, the set of model elements $V$, element influences forming the set $E$, and the set of element update rules $F$, comprise an *Executable Model*, $\mathcal{M}(V, E, F)$, a model that includes all the necessary information for simulation and dynamic analysis.

We use the Discrete, Stochastic, Heterogeneous simulator (DiSH) [36] which allows for simulations of discrete models with various types of update functions, and has several different simulation schemes, that can be either deterministic or stochastic. For the analysis we conducted here, we used the USB-RSQ simulation scheme in DiSH (uniform, step-based, random-order, sequential update scheme, described in detail in [36]). It has been shown previously [36, 37] that, by taking into account the randomness in timing of signaling events, the USB-RSQ simulation scheme is able to recapitulate the network dynamics within cells. DiSH simulates the models starting from an *initial state* $\boldsymbol{q}_{\mathcal{M},0} = (s_{v_1,0}, s_{v_2,0}, ..., s_{v_N,0})$ (assigned before simulations), where $s_{v_i,0}$ denotes the state value of element $v_i$ at time point $t = 0$, and for a pre-defined number of time steps, $T$ (e.g., when the steady state is reached). Each such simulation run, $r$, yields for every model element $v_i \in V$, a trajectory of values, $\boldsymbol{s}_{v_i}^r = (s_{v_i,1}^r, s_{v_i,2}^r, ... s_{v_i,T}^r)$, where $s_{v_i,t}^r$ is the state value of element $v_i$ at time point $t$ ($t = 1, ..., T$) within run $r$. Due to the randomness of the update scheme, element trajectories may vary across multiple runs that start with the same initial state. Therefore, for the same time step $t$, following the approach from [36], we compute the mean of values $s^r_{v_i,t}$ across different runs, to obtain average trajectories for all elements. More formally, we compute an *average element trajectory* of element $v_i$ as:

$$\bar{s}_{v_i} = \frac{1}{R}\sum_{r=1}^{R} s^r_{v_i} = \frac{1}{R}\sum_{r=1}^{R}\left(s^r_{v_i,1}, s^r_{v_i,2}, \ldots s^r_{v_i,T}\right)$$
$$= \left(\bar{s}_{v_i,1}, \bar{s}_{v_i,2}, \ldots \bar{s}_{v_i,T}\right) \qquad (3)$$

where $R$ is the overall number of conducted simulation runs. For example, in Figure 1 (B), we illustrate simulation trajectories for elements of the toy models in Figure 1 (A). We denote *average model state* for model $\mathcal{M}(V, E, F)$ at time step $t$ as a vector of average element states at time step $t$:

$$\boldsymbol{q}^{avg}_{\mathcal{M},t} = \left(\bar{s}_{v_1,t}, \bar{s}_{v_2,t}, \ldots, \bar{s}_{v_N,t}\right) \qquad (4)$$

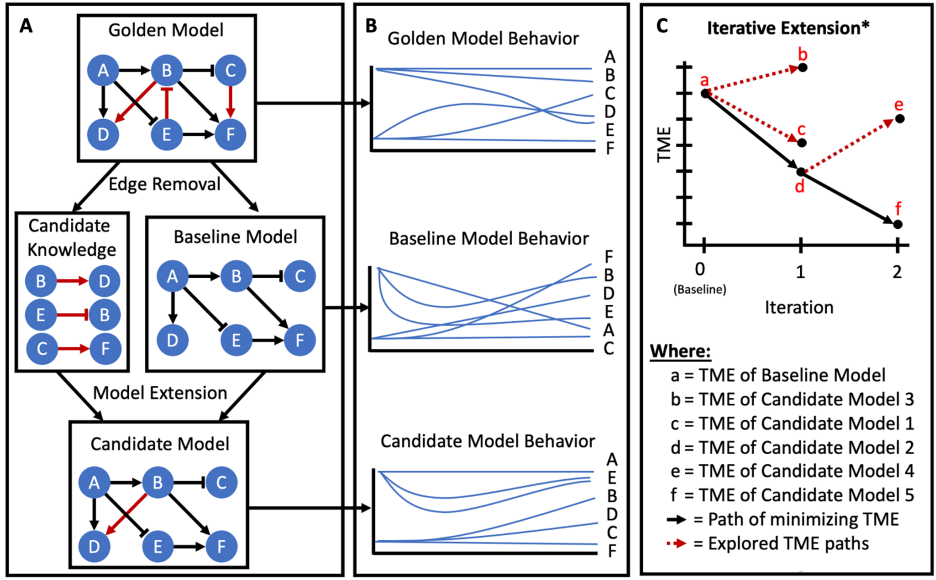

Figure 1. A toy example illustrating directed cyclic network models explored in this work and the flow of the proposed methodology for evaluating extension algorithms. (A) (top) An example Golden Model used in evaluation; (middle) Example input graphs, Candidate Knowledge, and Baseline Model, used in extension methods ([28-30] and this work); (bottom) An example Candidate Model recommended by extension methods. (B) Average element trajectories obtained from stochastic simulation for the three example models (Golden, Baseline, and Candidate). (C) An example iterative procedure that uses the Total Model Error (TME) metric to evaluate each intermediate Candidate Model.

We define *model behavior* resulting from a specific initial model state $\boldsymbol{q}_{\mathcal{M},0} = (S_1, S_2, \ldots, S_N)$ as:

$$\boldsymbol{Q}_{\mathcal{M}} = \left(\boldsymbol{q}_{\mathcal{M},0}, \boldsymbol{q}^{avg}_{\mathcal{M},1}, \ldots, \boldsymbol{q}^{avg}_{\mathcal{M},T}\right) \qquad (5)$$

## 2.2 Extension method inputs

We define here inputs used by extension methods and by our evaluation methodology: Baseline Model, Golden Model, and Candidate Knowledge.

Existing models of a system of interest are often leveraged and contextualized for a specific purpose. The Baseline Model is the existing, high confidence model before updating with extensions. As a special case, we can also assume that the Baseline Model is an empty network with no nodes or edges. The Golden Model is assumed to contain all relevant knowledge about the system, including accurate element relationships and update functions. The Candidate Knowledge is a set of directed edges, including their source and target nodes, which are candidates for addition to the Baseline Model.

Given the Golden Model knowledge, through simulations, for different initial states representing different conditions and scenarios, we can obtain Golden Model behavior, $\boldsymbol{Q}_{GM}$, as in Equation 5. $\boldsymbol{Q}_{GM}$ represents the true expected behavior of the system being modeled. As part of $\boldsymbol{Q}_{GM}$, we also obtain the average Golden Model state at the final simulation time step $T$ (e.g., steady state), $\boldsymbol{q}^{avg}_{GM,T}$.

The above definition of Golden Model is important for the rest of our discussion since Golden Model is used as an input to our evaluation methodology. However, in real scenarios, the Golden Model is usually not known in advance. Instead, the goal of model assembly and extension algorithms is to discover the Golden Model, while only the real system behavior, i.e., measured state values for system components, may be available. The system state data can be used to form the target behavior $\widehat{\boldsymbol{Q}}$. Ideally, the Golden Model behavior is identical to the target behavior, $\boldsymbol{Q}_{GM} = \widehat{\boldsymbol{Q}}$. The target state at time $T$ is part of the target behavior and is denoted as $\widehat{\boldsymbol{q}}_T$.

As will be detailed in the following sub-sections, extension algorithms start with the Baseline Model for which $\boldsymbol{Q}_{BM} \neq \widehat{\boldsymbol{Q}}$. Next, they add selected edges from the Candidate Knowledge to create new models, called Candidate Models, which are then iteratively updated and simulated to obtain $\boldsymbol{Q}_{CM}$ in each iteration, and to ultimately find a model that most closely reproduces the target behavior $\widehat{\boldsymbol{Q}}$.

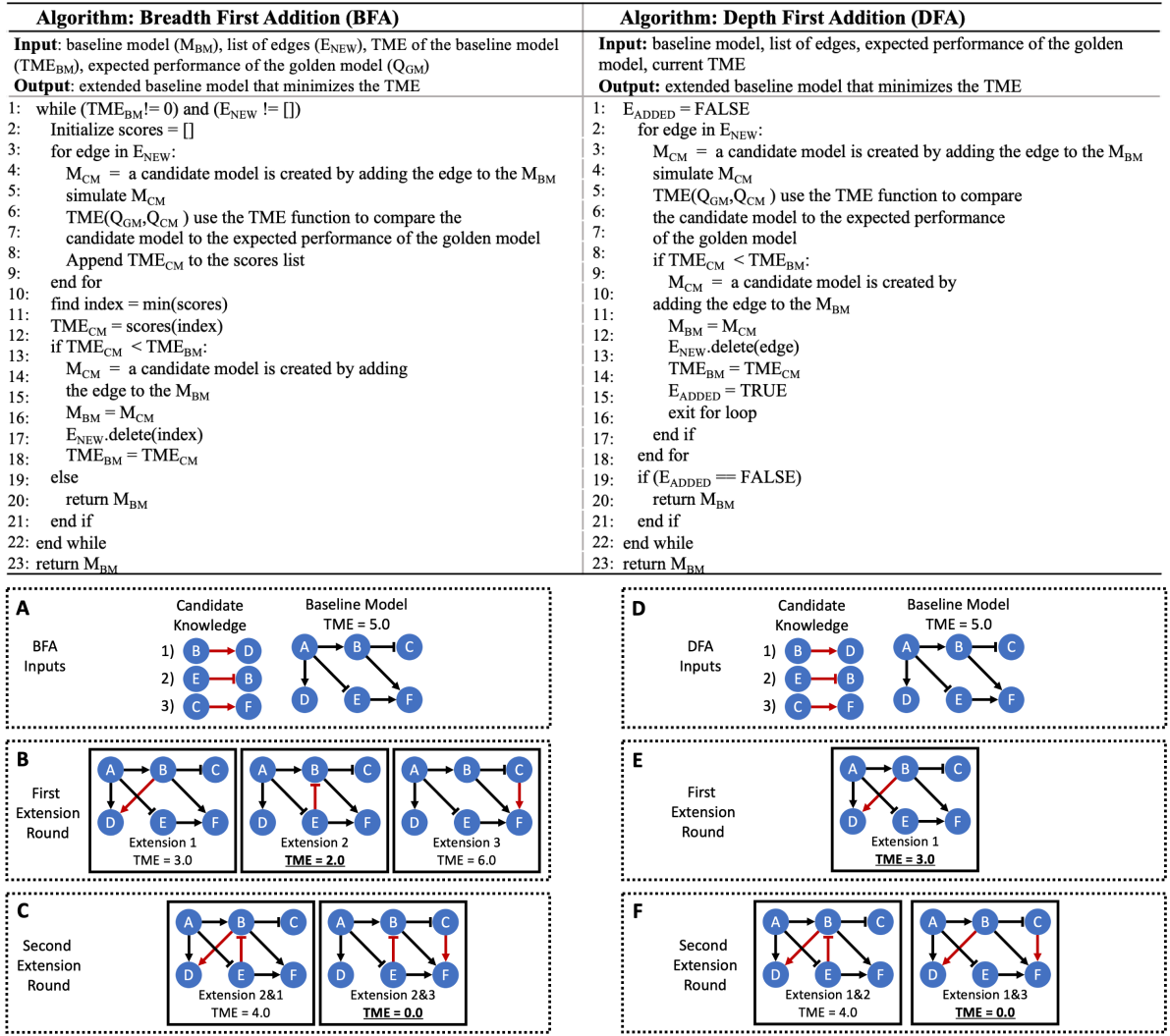

Figure 2. The Breadth and Depth First Addition (BFA and DFA, respectively) algorithms. <u>Top</u>: The pseudocode for the two algorithms. <u>Bottom</u>: An example illustrating the Candidate Knowledge and Baseline Model inputs and steps for BFA and DFA algorithms: (A, D) The inputs to the BFA and DFA algorithms. (B) In the BFA extension process, the Baseline Model is extended with single interactions from Candidate Knowledge and the TME is calculated for each Candidate Model. The Candidate Model with the lowest TME is selected and becomes the Baseline Model for the next iteration. (E) In the DFA extension process, the Baseline Model is extended with a single interaction from Candidate Knowledge and the TME is calculated to determine if the Candidate Model has a lower TME than the Baseline Model. As soon as the TME decreases, that edge of Candidate Knowledge is incorporated into the Candidate Model, and it becomes the Baseline Model for the next iteration. (C, F) For both algorithms, the process is repeated with the remaining Candidate Knowledge until all edges are added back, the TME reaches zero, or there are no edges that reduce the TME below its current lowest value.

### 2.3 Model evaluation metric

Given two models, $\mathcal{M}_1$ and $\mathcal{M}_2$, if they have the same element sets, $V_{\mathcal{M}_1} \equiv V_{\mathcal{M}_2} \equiv V$ ($N = |V|$), and if we simulate them starting from the same initial state, $\boldsymbol{q}_{\mathcal{M}_1,0} = \boldsymbol{q}_{\mathcal{M}_2,0} = (s_{v_1,0}, s_{v_2,0}, \dots, s_{v_N,0})$, to obtain their behaviors, $\boldsymbol{Q}_{\mathcal{M}_1}$ and $\boldsymbol{Q}_{\mathcal{M}_2}$, respectively, we can compute the difference between the two model behaviors, $\Delta_t(\boldsymbol{Q}_{\mathcal{M}_1}, \boldsymbol{Q}_{\mathcal{M}_2})$ at any simulation time step $t$ as:

$$\Delta_t(\boldsymbol{Q}_{\mathcal{M}_1}, \boldsymbol{Q}_{\mathcal{M}_2}) = \sum_{i=1}^{N} |\bar{s}_{v_i,t}^{\mathcal{M}_1} - \bar{s}_{v_i,t}^{\mathcal{M}_2}| \quad (6)$$

In other words, $\Delta_t$ finds the absolute difference between an element's average state in time step $t$, in model $\mathcal{M}_1$ ($\bar{s}_{v_i,t}^{\mathcal{M}_1}$) and in model $\mathcal{M}_2$ ($\bar{s}_{v_i,t}^{\mathcal{M}_2}$) and sums these differences across all model elements.

From (6), we derive the *Total Model Error (TME)* metric, as $\Delta_T$, when $t = T$, between a Candidate Model behavior $\boldsymbol{Q}_{CM}$ and known target behavior $\widehat{\boldsymbol{Q}}$:

$$\text{TME}(\boldsymbol{Q}_{CM}, \widehat{\boldsymbol{Q}}) = \Delta_T(\boldsymbol{Q}_{CM}, \widehat{\boldsymbol{Q}}) = \sum_{i=1}^{N} |\bar{s}_{v_i,T}^{CM} - \hat{s}_{v_i,T}| \quad (7)$$

Or, in the case when a Golden Model is used:

$$\text{TME}(\boldsymbol{Q}_{CM}, \boldsymbol{Q}_{GM}) = \Delta_T(\boldsymbol{Q}_{CM}, \boldsymbol{Q}_{GM}) = \sum_{i=1}^{N} |\bar{s}_{v_i,T}^{CM} - \bar{s}_{v_i,T}^{GM}| \quad (8)$$

Besides the above defined $\Delta_t$, other types of functions could be used to compute the difference between two models, such as the squared error, or more statistic-based evaluation methods like the Chi-squared test to compare the distribution of model states at time step $t$. We use the absolute difference of the model's end state ($t = T$) for a few reasons: it would not exaggerate the effect of large differences (as would be observed in the squared error); it is less computationally expensive than the Chi-squared test; and it more accurately matches how computational biologists compare computational model simulations against sparse biological measurements, where the full time-course of the model elements is often unknown.

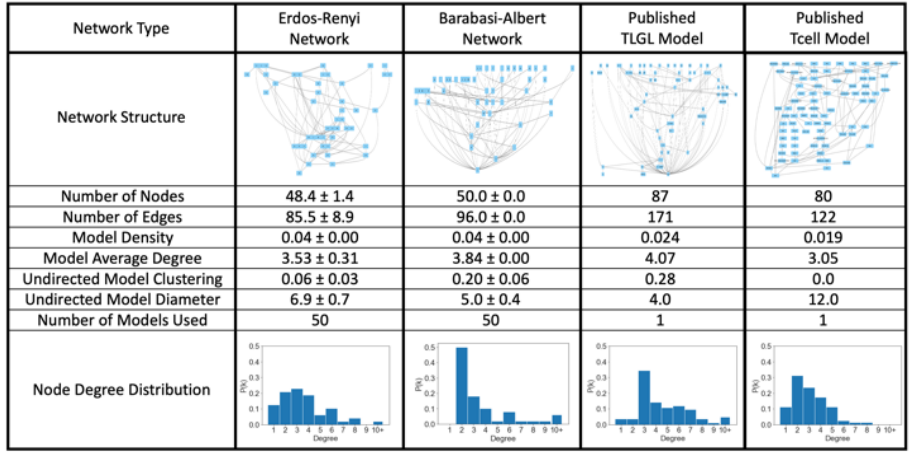

Figure 3. Network structure illustration, standard graph attributes, and node degree distribution histograms for different network types: Erdos-Renyi random networks, Barabasi-Albert scale-free networks, and two human-curated published biological networks, TLGL and Tcell.

### 2.4 Methodology for evaluating model extension

In this work, we are interested in evaluating automated model extension, that is, the limitations of automatically extending the Baseline Model with behavior $Q_{BM}$ to achieve the target or Golden Model behavior $Q_{GM}$. Therefore, in our studies we assume that the Golden Model is known, and to obtain Baseline Models we use the procedure illustrated in Figure 1 (A) and described as follows.

For a given Golden Model, we create multiple Baseline Models by *removing edges* from the Golden Model, in order to disrupt its behavior and to determine whether the extension algorithms are able to recover the Golden Model from a range of Baseline Models. The removed edges form the Candidate Knowledge sets (Figure 1 (A)). The extension algorithms are given the Baseline Model and the Candidate Knowledge and tasked with extending the Baseline Model using edges from the Candidate Knowledge, to create Candidate Models (Figure 1 (A)) and reproduce the Golden Model behavior.

Using the DiSH simulator, we simulate the Golden, Baseline, and Candidate Models to observe how elements of each model behave over time, and to obtain model behaviors $Q_{GM}$, $Q_{BM}$, $Q_{CM}$, respectively (Figure 1 (B)). The goal of this procedure is to find Candidate Model(s) with behavior similar to the Golden Model behavior. By tracking TME (Equation 8) across consecutive extension iterations, we can add Candidate Knowledge to the Baseline Model to form new Candidate Models and determine whether these new models perform more closely to the Golden Model (Equation 8). If the TME decreases, the Candidate Model is considered an improvement to the Baseline Model. If the TME increases, the Candidate Model is considered worse than the model from previous iteration, and the Candidate Knowledge incorporated is removed from the model. At each iteration, all Candidate Knowledge is added one interaction at a time and the TME is calculated. Candidate Knowledge with the largest decrease of TME is incorporated.

### 2.5 The Breadth First and Depth First Algorithms

In this analysis, we employ two algorithms to illustrate two different philosophies in automated assembly and extension; namely (i) incorporating the least amount of information necessary into the model that best improves the model and (ii) incorporating the most amount of information into the model as long as it relates to and improves the model. These algorithms are called the: (i) Breadth First Addition (BFA) algorithm that compares all potential additions against each other to only add the best supported information at any one time, and the (ii) Depth First Addition (DFA) algorithm that incorporates any new information that improves the model. The pseudocode for the two algorithms is shown in Figure 2 (top) and we depict example demonstrations for both algorithms in Figure 2 (bottom).

The Breadth First Addition (BFA) algorithm starts by evaluating the contribution of each new edge to decreasing TME, that is, it simulates the model that consists of the original Baseline Model and a selected new edge, and then computes TME of that extended model according to Eq 8. Next, it permanently incorporates the new edge that leads to the largest decrease in the original TME, and then it repeats the steps with this new extended model, i.e., similar to what was done with the original model, it evaluates addition of the remaining edges to this new model by computing their TME values. This process is repeated until at least one of the following conditions is satisfied: (i) the extended model matches the expected end values of the Golden Model; (ii) there are no more edges to evaluate; (iii) no edge can be added to the Baseline Model without increasing TME. The pseudocode and the toy example for the BFA algorithm are shown in **Error! Reference source not found.** (left).

The Depth First Addition (DFA) algorithm, similar to the BFA algorithm, starts with evaluation of edges by computing their contribution to decreasing TME of the Baseline Model. Different from BFA, as soon as it finds an edge which leads to a TME lower than the current TME, it adds that edge to the Baseline Model. These steps are then repeated using the new extended model and the remaining edges. Same as for the BFA algorithm, the DFA algorithm stops when at least one of the three conditions above, (i)-(iii) is satisfied. The pseudocode and the toy example for the DFA algorithm are shown in **Error! Reference source not found.** (right).

## 3 RESULTS

We describe here our experimental setup, including the set of benchmarks that we created (Sections 3.1 and 3.2), and we follow with a discussion of the outcomes of our study (Sections 3.3-3.5).

### 3.1 Benchmarks: Synthetic and Curated Models

In this analysis, we explore how the BFA and DFA algorithms affect automated assembly and extension of two types of synthetic networks and two manually curated published biological signaling pathway networks.

The Erdos-Renyi (ER) network type is considered a random graph and does not share many similarities to biological networks. The Barabasi-Albert (BA) network type is a scale-free network that has many shared characteristics with biological networks (most notably their node-degree distribution). Since we generated the ER and BA networks in a random manner, we created 50 models for each network type. We employed the python package, NetworkX [38] to create all synthetic networks.

The last two networks we used in our studies are the human-curated biological model of T cell large granular lymphocyte (TLGL) leukemia [34] and the biological model of naïve T cell differentiation (Tcell) [35]. The TLGL model has been used previously [39, 40] to perform structural and dynamic analysis in order to identify potential therapeutic targets, while the Tcell model was created to explore the control circuitry of naïve T cell differentiation [41][42].

In Figure 3, we show example networks illustrating different structure of these models. We also list several descriptive statistics for networks to demonstrate the similarities and differences between these network types. Model Density is the fraction of edges present over all possible edges between nodes. Model Average Degree is the sum of each node's degree across all model nodes (with degree being the number of edges that are incident to the node), divided by the number of nodes in the graph. Undirected Model Clustering [43] is a measure of the degree to which nodes in a graph tend to cluster together in groups of local triangles. Undirected Model Diameter is the maximum distance from any node in the network to any other node. In the last row in Figure 3, we provide histograms of the Node Degree Distribution metric. In the case of ER and BA networks, the histograms show average values for 50 generated models.

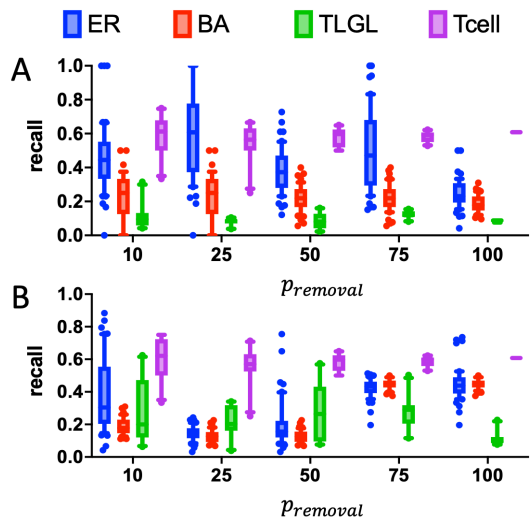

Figure 4. Recall distributions for all explored scenarios, for each network type (Erdos-Renyi - blue, Barabasi-Albert - red, TLGL - green, Tcell – purple) and at different edge removal probability ($p_{removal} \in [0.10, 0.25, 0.50, 0.75, 1.00]$). (A) BFA algorithm results and (B) DFA algorithm results.

### 3.2 Experimental Setup

For the purposes of the evaluation discussed here, we assume that each model element $v_i \in V$ ($i = 1..N$, where $N = |V|$), can be in one of the three states, OFF (value 0), LOW activity (value 1), and HIGH activity (value 2). This assumption makes the synthetic networks comparable to the published biological models. We randomly initialized the synthetic networks (as they are not based on human-curated or biological knowledge) while we initialized the Tcell [35] and TLGL [34] models based on the values listed in their corresponding publications. As nodes and edges are added back into the model, we assume that the initial state value of each model element $v_i$, is $s_{v_i,1}^r = 1$. For each created model, we conducted $R = 100$ simulation runs. For synthetic models, we simulated ER and BA models each with $T = 2{,}500$ time steps, while we simulated human curated models—TLGL and Tcell— for $T = 5{,}000$ time steps. The simulation length was governed by how long each network type required to reach a steady state.

### 3.3 Network structure and baseline information complicate model assembly

For each Golden Model, we used five different *removal probabilities* $p_{removal} \in [0.10, 0.25, 0.50, 0.75, 1.00]$ to randomly select edges for removal from the Golden Model. Edges that were removed formed the Candidate Knowledge and the remaining edges formed the Baseline Model. When $p_{removal} = 1.00$, the Baseline Model is empty (no edges) and both the BFA and DFA algorithms will attempt to reassemble the biological networks with only Candidate Knowledge. In all conducted studies ($p_{removal} \in [0.10, 0.25, 0.50, 0.75, 1.00]$), both the BFA and DFA algorithms were given the exact same Baseline Models and Candidate Knowledge and tasked to reconstruct the Golden Model. The recall—or ratio of edges returned to the Baseline Model out of all removed edges—is shown in Figure 4 for each network type (Erdos-Renyi - blue, Barabasi-

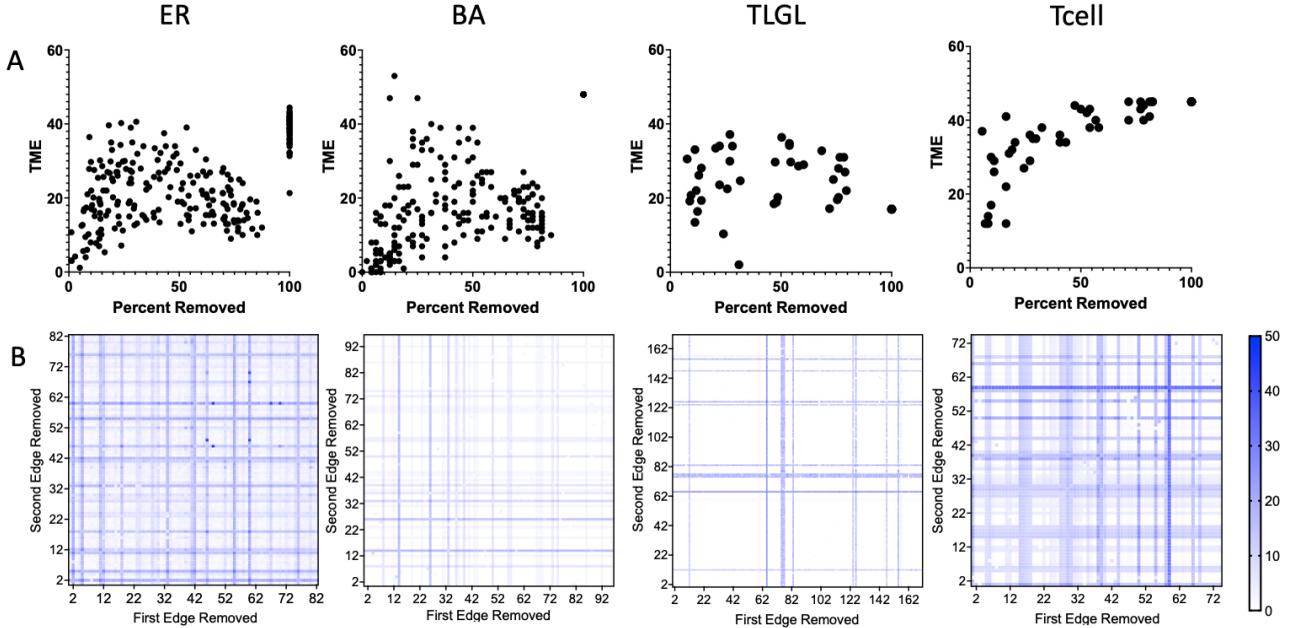

Figure 5. (A) Percent Removed plotted against TME for the ER, BA, TLGL, and Tcell network types. (B) The effect of the removal of pairs of edges from the network, first edge index indicated by the x-axis value, second edge index indicated by the y-axis value. The TME values are represented with shades of blue, from the minimum observed (i.e., no error, TME=0, shown in white) to the maximum observed (TME=50, shown in blue). Solid blue lines show the importance of particular edges to model performance and TME.

Albert - red, TLGL - green, Tcell - purple) and each algorithm (BFA – part A, DFA – part B).

In general, network type drastically affects recall rates, and for the most part, each network's recall trends down with higher $p_{removal}$. This makes intuitive sense as the more edges that are removed from each network, the more information there is to add back, and therefore the recall has a larger denominator (i.e., the size of the Candidate Knowledge set). Even with many missing edges, both BFA and DFA can still converge on local minima as long as each edge reduces TME. Both ER and Tcell network types correspond to higher rates of recall than in BA and TLGL. As both BFA and DFA add edges back based on each edge's effect on TME, this points to ER and Tcell networks having more edges which tangibly reduce TME. BA networks are noted for their hub and spoke structure, with a small number of highly connected nodes, and a large number of sparsely connected nodes. These networks are known for their redundancy, with the removal of an edge often compensated for by the rest of the network, the behavior that is observed in our results (Figure 4).

### 3.4 Model performance is difficult to encapsulate into one metric to optimize

We also examined the relationship between the selected $p_{removal}$ and TME. We expected the TME to be proportional to the amount of the information removed from the model (i.e., the number of edges in the Candidate Knowledge). To explore the effect of network structure on automated assembly and extension, we evaluated the starting TME of each Baseline Model. For each Baseline Model of each network type, we calculated the actual percentage of edges removed based on the $p_{removal}$. This percentage was termed the "Percent Removed". For each network type, we plotted the Percent Removed from the Golden Model and the TME before extension started. Next, starting with a Golden Model of each network type, we removed every combination of two edges and calculated the TME of the resultant Baseline Models. The results of these two analyses are shown in Figure 5.

We observed from our analysis that TME is not proportional to missing information and that the contribution of different edges to the model's TME can vary. At higher levels of Percent Removed, the relationship to TME is not linear. This points to the fact that even with only a few edges missing, a model can have quite high TME. We found that while TME does generally increase with more information removed, this increase is not directly proportional or consistent with information removed. TME functions as a simplified error function that approximates the Baseline Models deviation from Golden Model behavior but does not completely reflect how much information is missing from the Baseline Model or indicate how much information the algorithm must add back.

Additionally, network type has a large influence on the TME response to missing information. Networks like the Barabasi-Albert networks appear more robust to information removal, with no single edge resulting in large changes in TME. This same behavior is not observed in the Erdos-Renyi or human curated models where only a few edges can strongly affect TME. Indeed, returning to Figure 4, it appears that BA networks are some of the hardest to assemble and extend with automated methods relying on error evaluation, due to each edge only contributing a little to TME. A more comprehensive error function would require more information about the Golden Model's network structure and dynamics; however, this proves elusive as the more information about the Golden Model there is, the easier this problem becomes.

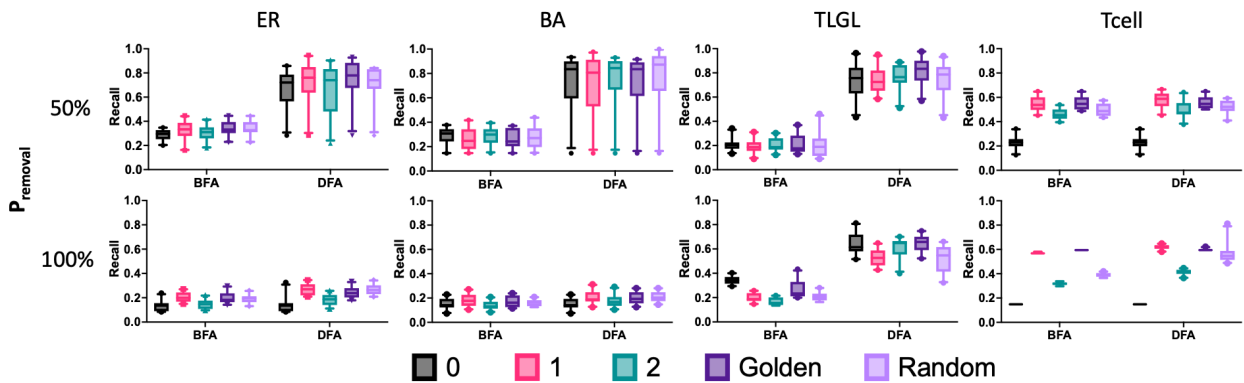

Figure 6. The ten networks for each network type were disassembled and then reassembled (either through BFA or DFA) under different initialization schemes. In "0" new model elements are initialized with a starting value of 0. Similarly, "1" and "2" follow similar schemes. "Golden" initializes the model element as it would be seen in the Golden Model, while "Random" randomly initializes the model element. In each assembly, the number of edges added back were recorded and used to calculate each assembly method's recall.

### 3.5 Initialization values play a small but important role in network assembly

Finally, when adding Candidate Knowledge back into the model, if a new node is introduced into the Baseline Model, there is no information surrounding how it should be initialized. In Figure 6 we show the effects of different initialization assumptions when adding Candidate Knowledge back into the model. Each network type was extended with BFA and DFA algorithms using one of five different initialization schemes: initializing new model elements with a fixed value (0, 1 or 2), initializing the model with the correct initialization used in the Golden Model, and randomly assigning an initial value. In general, initialization does not play a large role in automated assembly or extension. In Figure 6, there appears to be little difference between initialization types for the ER and BA network types, and the TLGL model. Although the human curated models (TLGL and Tcell) do diverge slightly from this trend, this is much more prominent for the Tcell model, which is an outlier, with automated model assembly and extension suffering due to the focused nature of the model. This is not to say that initialization is a problem that can be disregarded in model assembly, rather it is to be considered after the correct structure of the model has been identified. This is particularly true in the case of logical models where initialization can impact downstream model elements depending upon nature of the logic functions that are used. For example, initializing to 0 a model element involved in many "AND" operations will affect downstream model elements. As described in Section 2.1, we used summation functions in this analysis. This choice likely made the role of initialization less important, as the inclusion of a new edge (and thus, a new regulator for some element in the model) would not impact the effect of other regulators in such a substantial way as would be present with logic update functions.

Taken together, the discussion in Sections 3.3-3.5 and Figures 4-6 demonstrate the key difficulties to automated model assembly and extension. Several methods exist which create such automated pipelines but do not focus on how they incorporate biological information into executable models [27, 44, 45]. To date, only a few methods have been proposed to automatically assemble and extend models, while also evaluating the available information and its impact on the created executable model [28-30]. Still, even these methods do not fully assess the structural and dynamic impacts of adding new biological information to an executable model, and therefore do not address the complexities to this problem.

## 4 CONCLUSION

In this paper, we have presented an automated assembly and extension pipeline to depict the types and magnitudes of the problems facing computational and system biologists as they work to solve automated model assembly. Through the largest assembly and extension analysis of synthetic and human-curated models to date, we have characterized the complexities of the automated model assembly problem. Our findings demonstrate that iterative model assembly, devoid of context, lacking starting structural information in the form of a baseline model, and without robust dynamic information describing the golden model's behavior, is intractable. More often, model assembly creates models which perform similarly in dynamics, but do not represent the full information of a full "Golden" network.

In this paper, we have demonstrated that particular focus must be paid to a model's structure and baseline information, as these can complicate model assembly. In picking a metric to optimize during model assembly, we have illustrated that a single metric more often serves to simplify the golden model, rather than recapitulate it. Lastly, we have shown that initializing model elements only play a small role in network assembly.

In future work, we plan to further investigate the effect of network type, additional parametrization of update functions (e.g., timing effects), methods to determine initial state for simulations, and other error functions on the quality of recommended Candidate Models. We will also explore the effect of erroneous Candidate Knowledge on extension methods.

## Acknowledgment

NMZ is the corresponding author. This work was funded in part by DARPA award W911NF-17-1-0135. The authors would like to thank Kai-Wen Liang for his instrumental work in the implementation of the BFA algorithm.

## References

[1] J. M. Epstein, "Why Model?," 2008. [Online]. Available: http://jasss.soc.surrey.ac.uk/11/4/12.html.

[2] E. A. Sobie, Y.-S. Lee, S. L. Jenkins, and R. Iyengar, "Systems biology—biomedical modeling," *Sci. Signal.,* vol. 4, no. 190, pp. tr2-tr2, 2011.

[3] J. Schäfer and K. Strimmer, "An empirical Bayes approach to inferring large-scale gene association networks," *Bioinformatics,* vol. 21, no. 6, pp. 754-764, 2004.

[4] R. Küffner, T. Petri, P. Tavakkolkhah, L. Windhager, and R. Zimmer, "Inferring gene regulatory networks by ANOVA," *Bioinformatics,* vol. 28, no. 10, pp. 1376-1382, 2012.

[5] K. Raza, "Fuzzy logic based approaches for gene regulatory network inference," *Artificial intelligence in medicine,* 2018.

[6] P. B. Madhamshettiwar, S. R. Maetschke, M. J. Davis, A. Reverter, and M. A. Ragan, "Gene regulatory network inference: evaluation and application to ovarian cancer allows the prioritization of drug targets," *Genome medicine,* vol. 4, no. 5, p. 41, 2012.

[7] P. D'haeseleer, S. Liang, and R. Somogyi, "Genetic network inference: from co-expression clustering to reverse engineering," *Bioinformatics,* vol. 16, no. 8, pp. 707-726, 2000.

[8] J. Linde, S. Schulze, S. G. Henkel, and R. Guthke, "Data-and knowledge-based modeling of gene regulatory networks: an update," *EXCLI journal,* vol. 14, p. 346, 2015.

[9] N. Wani and K. Raza, "Integrative Approaches to Reconstruct Regulatory Networks From Multi-Omics Data: A Review of State-of-the-Art Methods," 2018.

[10] M. Hecker, S. Lambeck, S. Toepfer, E. Van Someren, and R. Guthke, "Gene regulatory network inference: data integration in dynamic models—a review," *Biosystems,* vol. 96, no. 1, pp. 86-103, 2009.

[11] M. Banf and S. Y. Rhee, "Enhancing gene regulatory network inference through data integration with markov random fields," *Scientific reports,* vol. 7, p. 41174, 2017.

[12] M. Recamonde-Mendoza, A. V. Werhli, and A. Biolo, "Systems biology approach identifies key regulators and the interplay between miRNAs and transcription factors for pathological cardiac hypertrophy," *Gene,* Mar 4 2019, doi: 10.1016/j.gene.2019.02.056.

[13] A. Fabregat *et al.*, "The Reactome Pathway Knowledgebase," *Nucleic Acids Research,* vol. 46, no. D1, pp. D649-D655, 2018, doi: 10.1093/nar/gkx1132.

[14] P. D. Karp, M. Riley, S. M. Paley, and A. Pellegrini-Toole, "The MetaCyc Database," *Nucleic acids research,* vol. 30, no. 1, pp. 59-61, 2002. [Online]. Available: http://www.ncbi.nlm.nih.gov/pubmed/11752254 http://www.pubmedcentral.nih.gov/articlerender.fcgi?artid=PMC99148.

[15] D. Türei, T. Korcsmáros, and J. Saez-Rodriguez, "OmniPath: guidelines and gateway for literature-curated signaling pathway resources," *Nature Methods,* vol. 13, no. 12, pp. 966-967, 2016, doi: 10.1038/nmeth.4077.

[16] D. Szklarczyk *et al.*, "The STRING database in 2017: quality-controlled protein-protein association networks, made broadly accessible," *Nucleic acids research,* vol. 45, no. D1, pp. D362-D368, 2017, doi: 10.1093/nar/gkw937.

[17] C. F. Schaefer *et al.*, "PID: the pathway interaction database," *Nucleic Acid Res,* vol. 37, 2009, doi: 10.1093/nar/gkn653.

[18] D. N. Slenter *et al.*, "WikiPathways: a multifaceted pathway database bridging metabolomics to other omics research," *Nucleic acids research,* vol. 46, no. D1, pp. D661-D667, 2018, doi: 10.1093/nar/gkx1064.

[19] V. Chelliah *et al.*, "BioModels: ten-year anniversary," *Nucleic Acids Research,* vol. 43, no. D1, pp. D542-D548, 2015, doi: 10.1093/nar/gku1181.

[20] T. Helikar *et al.*, "The Cell Collective: toward an open and collaborative approach to systems biology," *BMC Syst Biol,* vol. 6, 2012, doi: 10.1186/1752-0509-6-96.

[21] M. Kanehisa, M. Furumichi, M. Tanabe, Y. Sato, and K. Morishima, "KEGG: new perspectives on genomes, pathways, diseases and drugs," *Nucleic Acids Research,* vol. 45, no. D1, pp. D353-D361, 2017, doi: 10.1093/nar/gkw1092.

[22] M. A. Valenzuela-Escárcega, G. Hahn-Powell, and M. Surdeanu, "Description of the Odin Event Extraction Framework and Rule Language," 2015. [Online]. Available: http://arxiv.org/abs/1509.07513.

[23] G. Ferguson and J. F. Allen, "TRIPS: An integrated intelligent problem-solving assistant," 1998: AAAI Press, pp. 567-572, doi: 10.1080/00021369.1971.10860128. [Online]. Available: https://dl.acm.org/citation.cfm?id=295737 http://dblp.uni-trier.de/db/conf/aaai/aaai98.html#FergusonA98%5Cnhttp://www.aaai.org/Papers/AAAI/1998/AAAI98-080.pdf

[24] K. Hakala, S. Van Landeghem, T. Salakoski, Y. Van de Peer, and F. Ginter, "Application of the EVEX resource to event extraction and network construction: Shared Task entry and result analysis," *BMC Bioinformatics,* vol. 16, no. Suppl 16, pp. S3-S3, 2015, doi: 10.1186/1471-2105-16-S16-S3.

[25] F. Buchel *et al.*, "Path2Models: large-scale generation of computational models from biochemical pathway maps," *BMC Syst Biol,* vol. 7, no. 1, p. 116, 2013, doi: 10.1186/1752-0509-7-116.

[26] B. M. Gyori, J. A. Bachman, K. Subramanian, J. L. Muhlich, L. Galescu, and P. K. Sorger, "From word models to executable models of signaling networks using automated assembly," *Molecular systems biology,* vol. 13, no. 11, pp. 954-954, 2017, doi: 10.15252/msb.20177651.

[27] R. Sharp *et al.*, "Eidos, INDRA, & Delphi: From free text to

executable causal models," in *Proceedings of the 2019 Conference of the North American Chapter of the Association for Computational Linguistics (Demonstrations)*, 2019, pp. 42-47.

[28] K.-W. Liang, Q. Wang, C. Telmer, D. Ravichandran, P. Spirtes, and N. Miskov-Zivanov, "Methods to Expand Cell Signaling Models Using Automated Reading and Model Checking," Springer, Cham, 2017, pp. 145-159.

[29] Y. Ahmed, C. Telmer, and N. Miskov-Zivanov, "ACCORDION: Clustering and Selecting Relevant Data for Guided Network Extension and Query Answering," *arXiv preprint arXiv:2002.05748*, 2020.

[30] K. Sayed, K. N. Bocan, and N. Miskov-Zivanov, "Automated Extension of Cell Signaling Models with Genetic Algorithm," 2018/07//: IEEE, pp. 5030-5033, doi: 10.1109/EMBC.2018.8513431. [Online]. Available: https://ieeexplore.ieee.org/document/8513431/

[31] D. C. Kozen, "Depth-First and Breadth-First Search," in *The Design and Analysis of Algorithms*. New York, NY: Springer New York, 1992, pp. 19-24.

[32] P. Erdős and A. Rényi, "ON THE EVOLUTION OF RANDOM GRAPHS." [Online]. Available: http://leonidzhukov.net/hse/2014/socialnetworks/papers/erdos-1960-10.pdf.

[33] A.-L. Barabasi and R. Albert, "Emergence of scaling in random networks," *Science (New York, N.Y.)*, vol. 286, no. 5439, pp. 509-12, 1999, doi: 10.1126/SCIENCE.286.5439.509.

[34] R. Zhang et al., "Network model of survival signaling in large granular lymphocyte leukemia," *Proc Natl Acad Sci U S A*, vol. 105, no. 42, pp. 16308-13, Oct 21 2008, doi: 10.1073/pnas.0806447105.

[35] N. Miskov-Zivanov, M. S. Turner, L. P. Kane, P. A. Morel, and J. R. Faeder, "The duration of T cell stimulation is a critical determinant of cell fate and plasticity," (in eng), *Science signaling*, Research Support, N.I.H., Extramural Research Support, Non-U.S. Gov't Research Support, U.S. Gov't, Non-P.H.S. vol. 6, no. 300, p. ra97, Nov 5 2013, doi: 10.1126/scisignal.2004217.

[36] K. Sayed, Y.-H. Kuo, A. Kulkarni, and N. Miskov-Zivanov, "Dish simulator: capturing dynamics of cellular signaling with heterogeneous knowledge," presented at the Proceedings of the 2017 Winter Simulation Conference, Las Vegas, Nevada, 2017. https://github.com/pitt-miskov-zivanov-lab/dyse_wm

[37] S. M. Assmann and R. Albert, "Discrete Dynamic Modeling with Asynchronous Update, or How to Model Complex Systems in the Absence of Quantitative Information," in *Plant Systems Biology*, D. A. Belostotsky Ed. Totowa, NJ: Humana Press, 2009, pp. 207-225.

[38] *Proceedings of the Python in Science Conference (SciPy): Exploring Network Structure, Dynamics, and Function using NetworkX*. (2008). [Online]. Available: http://conference.scipy.org/proceedings/scipy2008/paper_2/

[39] Y. Ahmed, C. A. Telmer, and N. Miskov-Zivanov, "CLARINET: Efficient learning of dynamic network models from literature," *Bioinformatics Advances*, vol. 1, no. 1, p. vbab006, 2021.

[40] A. Saadatpour et al., "Dynamical and structural analysis of a T cell survival network identifies novel candidate therapeutic targets for large granular lymphocyte leukemia," *PLoS computational biology*, vol. 7, no. 11, p. e1002267, 2011.

[41] W. F. Hawse et al., "Cutting edge: differential regulation of PTEN by TCR, Akt, and FoxO1 controls CD4+ T cell fate decisions," *The Journal of Immunology*, vol. 194, no. 10, pp. 4615-4619, 2015.

[42] N. Miskov-Zivanov, M. Turner, L. Kane, P. Morel, and J. Faeder, "Model Predicts Duration of T Cell Stimulation is a Critical Determinant of Cell Fate and Plasticity, under submission," 2013.

[43] J. Saramäki, M. Kivelä, J.-P. Onnela, K. Kaski, and J. Kertesz, "Generalizations of the clustering coefficient to weighted complex networks," *Physical Review E*, vol. 75, no. 2, p. 027105, 2007.

[44] F. Büchel et al., "Path2Models: large-scale generation of computational models from biochemical pathway maps," *BMC Systems Biology*, vol. 7, no. 1, pp. 116-116, 2013, doi: 10.1186/1752-0509-7-116.

[45] B. M. Gyori, J. A. Bachman, K. Subramanian, J. L. Muhlich, L. Galescu, and P. K. Sorger, "From word models to executable models of signaling networks using automated assembly," *Molecular systems biology*, vol. 13, no. 11, 2017.

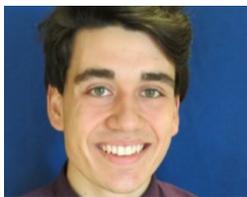

**Adam A. Butchy**. Adam is a PhD candidate in the Bioengineering Department at the University of Pittsburgh. Adam completed a B.S. in Chemical Engineering and a B.S. in Biochemistry at Villanova University. He is working on Discrete Modeling of Macrophage Activation and its role in the cancer microenvironment and lung.

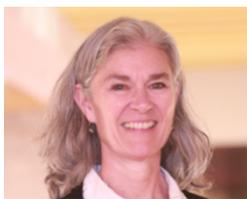

**Cheryl A. Telmer** Dr. Telmer is a Research Biologist at Carnegie Mellon University. Cheryl and Natasa began working together as iGEM advisors in 2013 and have expanded their collaborations through the DARPA Big Mechanism and World Modelers programs. Biologists are constantly trying new tools that have the potential to improve our understanding of complex systems, and the standardized representation and computational modeling approaches being developed by the Melody Lab are a great contribution.

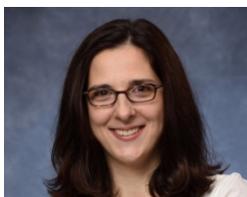

**Natasa Miskov-Zivanov** Dr. Miskov-Zivanov is an Assistant Professor of Electrical and Computer Engineering, Bioengineering, and Computational and Systems Biology at the University of Pittsburgh. She received a B.Sc. degree in electrical engineering and computer science from University of Novi Sad, Serbia and M.Sc. and Ph.D. degrees in electrical and computer engineering from Carnegie Mellon University. Before joining University of Pittsburgh as a faculty, she spent several years as a postdoctoral researcher in Computational and Systems Biology at the University of Pittsburgh, and as research scientist and instructor in Computer Science and in Electrical and Computer Engineering at Carnegie Mellon University. Dr. Miskov-Zivanov's research interests include hybrid, knowledge-driven and data-driven, model recommendation and reasoning for complex systems with applications in systems and synthetic biology.